# Landmark detection in Cardiac Magnetic Resonance Imaging Using A Convolutional Neural Network

Hui Xue[1], Jessica Artico[2], Marianna Fontana[3], James C Moon[2], Rhodri H Davies[2], Peter Kellman[1]

1. National Heart, Lung and Blood Institute, National Institutes of Health, Bethesda, MD, USA
2. Barts Heart Centre, Barts Health NHS Trust, London, UK
3. National Amyloidosis Centre, Royal Free Hospital, London, UK

## Corresponding author:

Hui Xue

National Heart, Lung and Blood Institute
National Institutes of Health
10 Center Drive, Bethesda
MD 20892
USA

Email: hui.xue@nih.gov

**Word Count: 3,074**

# Landmark detection in Cardiac Magnetic Resonance Imaging Using A Convolutional Neural Network

Hui Xue[1], Jessica Artico[2], Marianna Fontana[3], James C Moon[2], Rhodri Davies[2], Peter Kellman[1]

1. National Heart, Lung and Blood Institute, National Institutes of Health, Bethesda, MD, USA
2. Barts Heart Centre, Barts Health NHS Trust, London, UK
3. National Amyloidosis Centre, Royal Free Hospital, London, UK

**Key points:**

1. Proposed and validated a convolutional neural network solution for robust landmark detection for the long and short-axis views in CMR cine, LGE and T1 mapping and integrated an automated inline implementation on the MR scanner.

2. The large training set includes included more than 2,000 patients and 34,000 images, from two hospitals and the hold-out independent test set included more than 500 patients and 7,000 images.

3. High detection rate (96.6% to 99.8%) was achieved on the test set and comparison of RV insertion angle and LV length measured with models and manual labelling showed no significant differences between expert and AI on all imaging sequences.

**Summary statement:**

This study developed, validated and deployed a CNN solution for robust landmark detection in long and short-axis CMR images for cine, LGE and T1 mapping, with the performance comparable to the manual labelling.

**Abbreviations**

CMR = cardiac magnetic resonance, AHA = American Heart Association, CNN = convolutional neural network, CPU = central processing unit, LV = left ventricular, RV = right ventricular, RVI = right ventricular insertion, MOLLI = modified Look-Locker inversion recovery, LAX = long-axis, SAX = short-axis

**Abstract**

**Purpose**

To develop a convolutional neural network (CNN) solution for robust landmark detection in cardiac MR images.

**Methods**

This retrospective study included cine, LGE and T1 mapping scans from two hospitals. The training set included 2,329 patients and 34,019 images. A hold-out test set included 531 patients and 7,723 images. CNN models were developed to detect two mitral valve plane and apical points on long-axis (LAX) images. On short-axis (SAX) images, anterior and posterior RV insertion points and LV center were detected. Model outputs were compared to manual labels by two operators for accuracy with a t-test for statistical significance. The trained model was deployed to MR scanners.

**Results**

For the LAX images, success detection was 99.8% for cine, 99.4% for LGE. For the SAX, success rate was 96.6%, 97.6% and 98.9% for cine, LGE and T1-mapping. The L2 distances between model and manual labels were 2 to 3.5 mm, indicating close agreement between model landmarks to manual labels. No significant differences were found for the anterior RV insertion angle and LV length by the models and operators for all views and imaging sequences. Model inference on MR scanner took 610ms/5.6s on GPU/CPU, respectively, for a typical cardiac cine series.

**Conclusions**

This study developed, validated and deployed a CNN solution for robust landmark detection in both long and short-axis CMR images for cine, LGE and T1 mapping sequences, with the accuracy comparable to the inter-operator variation.

# Introduction

Cardiac magnetic resonance imaging (CMR) is emerging as a main-stream modality to image the cardiovascular system for diagnosis and intervention. Today's CMR imaging has advanced beyond the scope of imaging anatomy and can provide comprehensive quantitative measures of the myocardium, such as relaxometry T1, T2, and T2* (1) (2) to assess fibrosis, edema, and iron, tissue composition such as fat fraction (3) and physiology such as myocardial perfusion (4,5) and blood volume (6) mapping. These capabilities open new opportunities and simultaneously place new demands on image analysis and reporting. A fully automated solution brings increased objectivity, reproducibility and higher patient throughputs.

While imaging technology has greatly advanced, automated analysis and reporting of CMR is still catching up. In clinical practice, manual delineation by cardiologists remains the main approach to quantify cardiac function, viability and tissue properties (7). A recent study showed a detailed manual analysis can take 9 to 19 mins of an expert's time (8).

With the introduction of deep learning, in particular convolutional neural networks (CNN), recent studies have begun to report automated CMR analysis solutions with performance and robustness close to expert reading. Cardiac cine images can be automatically analyzed using CNN to measure ejection fraction and other parameters to match the expert level performance (9) and have demonstrated improved reproducibility in a multi-center trial (8). Cardiac perfusion images may be successfully analyzed and reported on MR scanners (10) using CNN and which has been shown to be an independent predictor of adverse cardiovascular outcomes (11). Current solutions have focused on automating the time-consuming step of segmenting the myocardium.

To achieve automated analysis and reporting of CMR, key landmark points must be located on the cardiac images. For example, right ventricular (RV) insertion points are needed to report quantitative maps using the standard AHA sector model (7). For the long-axis views, ventricular length can be measured if valve and apical points can be delineated. The variation of LV length is a useful biomarker and shown to be the principal component of left ventricular pumping in patients with chronic myocardial infarction (12). Furthermore, cardiac landmark detection can be useful on its own for applications such as automated imaging slice planning.

In this study we propose a generic CNN based solution for automatic cardiac landmark detection for CMR images. This solution is capable of detecting landmarks on both the long-axis (two-chamber CH2, three-chamber CH3 and four-chamber CH4) and short-axis (SAX) stacks. As illustrated in Figure 1, for every SAX view, the anterior and inferior RV insertion (A-RVI and I-RVI) and LV center points (C-LV) are detected. For the two-chamber view, the anterior and inferior (A-P and I-P) points are detected. The inferoseptal and anterolateral points (IS-P, AL-P) are detected for CH4. The inferolateral and anteroseptal points (IL-P, AS-P) are detected for CH3. The apical point (APEX) are detected for all LAX views. Trained CNN models were tested on cardiac cine, late gadolinium enhancement (LGE) and T1 maps derived from a modified Look-Locker inversion recovery (MOLLI) imaging sequence (1,13).

A large dataset was curated from two hospitals and split for training (N= 2,329 patients; 34,019 images) and for testing (N=531 patients; 7,723 images, no overlap with training set), including all three tested imaging sequences. The performance of the trained CNNs was quantitatively evaluated by comparing against manual labels for success rate and computing L2 distance between manual and model derived landmarks. To evaluate the feasibility of models for CMR reporting, two measures were computed from AI landmarks and compared to manual values: the angle of anterior RVI point and length of the LV. To demonstrate clinical feasibility, the trained CNN models were integrated on MR scanners using Gadgetron InlineAI (14) and used to automatically measure the LV length from long-axis cine imaging acquisition.

## Methods

*Data Collection*

In this retrospective study, a dataset was assembled from two hospitals (Barts Heart Centre, BHC; Royal Free Hospital, RFH). All cine and LGE scans were performed at the BHC and all T1 MOLLI images were acquired from the RFH. Both LAX and SAX views were acquired for cine and LGE and T1 mapping acquired 1-3 SAX slices per patient.

Table 1 summarizes the training and test datasets. For training, a total of 34,019 images were included for N=2,329 patients, with 29,214 cine and 3,798 LGE and 1,077 T1 images. Cine data was curated from three scan periods. All patients with LGE scans also had cine imaging. Data acquisition in every scan period was consecutive. The test set consisted of 7,723 images from 531 patients and were consecutively. There was no overlap between training and test sets.

Datasets were acquired using both 1.5 T (four MAGNETOM Aera, Siemens AG Healthcare, Erlangen, Germany) and 3 T (one MAGNETOM Prisma, Siemens AG Healthcare) MR scanners. In the training set, 1,790 patients were scanned with 1.5T scanners and 539 were scanned with 3T. In the test set, 462 patients used 1.5T MRI and 69 used 3T. Typically 30 cardiac phases were reconstructed for each heartbeat for every cine scan. For the training and testing purpose, the first phase (typically end-diastolic) and the end-systolic phase were selected. The rationale is that with large number of subjects, the cardiac phases will represent a sufficiently broad variation.

Data was acquired with the required ethical and/or secondary audit use approvals or guidelines (as per each center) that permitted retrospective analysis of anonymized data for the purpose of technical development, protocol optimization and quality control. All data was anonymized and delinked for analysis with approval by the local Office of Human Subjects Research (Exemption #13156).

*Imaging Sequences*

Cine imaging utilized a standard balanced SSFP sequence with typical imaging parameters: TR = 2.7/TE = 1.2 ms for 256 ×144 matrix, flip angle 40°, typical FOV 360 × 270 mm$^2$, slice thickness 8 mm with a gap of 2 mm, bandwidth 977 Hz/pixel. Cine acquisitions were performed with retrospective ECG gating (30 cardiac phases were reconstructed) and 2-fold parallel imaging acceleration using GRAPPA (15). For the SAX acquisition, a scan typically had 8 to 14 slices to cover the LV.

Phase sensitive inversion recovery (PSIR) LGE imaging was performed with a free-breathing sequence (16) for whole LV coverage with respiratory motion correction and averaging. Typical imaging parameters were: TR = 2.76/TE = 1.1, 256 × 144 matrix, flip angle 50°, typical FOV 360×270 mm$^2$, slice thickness 8 mm with a gap of 2 mm, single-shot BSSFP readout, 8 measurements per slice with 4 averages. The phase sensitive LGE reconstruction (17) was used to achieve insensitivity to inversion time. Previous studies (18) showed this free-breathing technique is more robust against respiratory motion and delivered improved LGE image quality.

T1 mapping used in this study used a previously published MOLLI protocol (1). Typical imaging parameters were: FOV 360×270 mm$^2$, 256x144 matrix size, 1085 Hz/pixel bandwidth, 35 degrees flip angle, and sampling strategy was 5s(3s)3s for pre-contrast T1 scans and 4s(1)3s(1s)2s for post-contrast scans. A retrospective motion correction algorithm (19) was applied to MOLLI images and then went through the T1 fitting (20) to estimate per-pixel maps.

*Data Preparation and Labeling*

Since the acquired field-of-view may vary between patients, all images were first resampled to a fixed 1mm$^2$ pixel spacing and padded/cropped to 400×400 pixels before input into the CNN. The cine MR imaging often causes a shadow across the FOV (Figure 2), as the tissue which is further away from receive coils on the chest and spine will have reduced signal intensity due to inhomogeneity of the surface coil receive sensitivity. To compensate for this shading, for every cine image in the dataset, a surface coil inhomogeneity correction algorithm (21) was applied to estimate slowly varying surface coil sensitivity which was used to correct this inhomogeneity . During training, either the original cine image or the corrected one was fed into the network with a probability p=0.5 to pick original version. This served as a data augmentation step. Other data augmentation used included adding random gaussian noise (prob. to add noise is 0.5, noise sigma was uniformly picked from 10-30% of the mean image value and adding blurring with a Gaussian kernel applied randomly to images (prob. to apply filtering p=0.5, filter sigma was uniformly picked from [0.5, 1.0, 2.0] pixel).

One operator (E1, 9 years of experiences) manually labelled all images for training and test (41742 images). A second operator (E2, 3 years of experience) was invited to label part of the test dataset to assess inter-operator variation. E2 labelled 1,100 images (Cine and LGE: 100 images for every LAX view, 200 images for SAX; T1 maps: 100 images). The VIA Image Annotator software (*http://www.robots.ox.ac.uk/~vgg/software/via/*) was used by both operators for manual labelling of landmarks. The data labeling took ~150 hours in total.

The SAX cine training set included 702 full stacks and 16,502 images for two cardiac phases. 3,803 images were acquired outside the LV and, therefore, contained no landmarks. The SAX cine test set contained 128 full stacks and 3,008 images with 813 images having no landmarks. The SAX LGE training set included 178 full stacks and 2,018 images, with 371 images having no landmarks. The SAX LGE test set contained 96 full stacks and 1,082 images with 222 images containing no landmarks.

*Model and Training*

The landmark detection problem was formulated as a "heat map" (22). As shown in Figure 3, every landmark point is convolved with a Gaussian kernel and the resulting blurred distribution represents the spatial probability of this landmark. Detecting three landmarks was equivalent to a semantic segmentation problem for four classes (background class and one object class for each landmark).

A variation of U-net architecture was implemented (23,24) for heat-map detection. As shown in Figure 4, the network was organized as layers for different spatial resolution. Each layer can contain several blocks. Each block had two convolution layers with batch normalization (25) and LeakyRelu activation functions (26). The network can be made deeper by inserting more resolution layers or by inserting more blocks. Going down the downsampling branch, the image spatial resolution was reduced by ×2 for every layer with the number of filters increased. Going up the upsampling branch, the spatial resolution was restored with a reduced number of filters. All convolution layers had filter size 3×3 with stride 1 and padding 1. The final convolution layer outputs a per-pixel score tensor which is converted to a probability tensor using a SoftMax. Figure 3 plots the specific network configuration used in this experiment.

In the data preparation step, all images were resampled and cropped to 400×400 pixels square. The CNN output score tensor had dimensions 400×400×4 (Figure 4). To train the network, the KL divergence was computed between ground-truth heat-map and SoftMax tensor of scores. Besides this entropy-based loss, the shape loss was further computed as the soft Dice ratio (22). The final loss was a sum of entropy-based loss and soft Dice ratio.

Two models were trained for LAX and SAX. For the LAX, all views were trained together as a multi-task learning task. Since the number of images for each LAX view was roughly equal, no extra data rebalancing strategy was applied. Instead, every minibatch randomly selected from CH2, CH3 or CH4 images and refined network weights.

The data for training was split with 90% of all patients for train and 10% for validation. The split was based on studies, so there was no data mixing across patients. The Adam optimizer was used with an initial learning rate of 0.001, betas were 0.9 and 0.999 and epsilon was 1e-8. Learning rate was reduced by x2 whenever the cost function plateaued. Training lasted 50 epochs (~4 hours) and the best model was selected as the one giving the best performance on the validation set. The CNN model was implemented using PyTorch (27) and training was performed on an Ubuntu 20.04 PC with four NVIDIA GTX 2080Ti GPU cards, each with 11GB RAM. Data parallelization was used across multiple GPU cards to speedup training.

Since there are more cine images than the other two categories, a fine-tuning strategy was implemented using transfer learning. For both LAX and SAX, a model was first trained with cine dataset and then fine-tuned with either LGE or T1 training sets. To perform the fine tuning, the initial learning rate was set to be 0.0005 and a total of 10 epochs were trained.

*Performance Evaluation and Statistical Analysis*

The trained model was applied to all test samples. All results were first visually reviewed to determine whether landmarks were missed or unnecessarily detected. For example, if a mid-SAX slice was marked as three landmark points (see Figure 1) and only two points were detected by model, this case was reported as a failed detection case. The detection rate was computed as the percentage of samples whose landmarks were correctly detected. For all samples with successful detection, the Euclidian distance between detected landmarks and labels was computed and reported separately for different slice views and different landmark points. Results from model detection and manual labels were compared and Euclidian distance between two operators were reported.

The detected key points were further processed to compute two derived measurements: a) the angle of anterior RV insertion point to LV center for SAX views; b) the length of LV for LAX views, computed as length from detected apical point to the middle point of two valve points (28). The model derived results were compared between manual labels. The results of the 1$^{st}$ operator were compared to the 2$^{nd}$ operator to give references for inter-operator variation.

Results were presented as mean ± standard deviation. T-test was performed and a *P*-value less than .05 was considered statistically significant (Matlab R2017b, Mathworks Inc., MA, USA).

*Model Deployment*

To demonstrate the clinical relevance of CMR landmark detection, an inline application was developed to measure LV length from LAX cine images automatically on the MR scanner. The trained LAX model was integrated onto MR scanners using the Gadgetron InlineAI toolbox (14). While the imaging was ongoing, the trained model was loaded and after the cine images were reconstructed, the model was applied to the acquired images as part of the image reconstruction workflow (inline processing) at the time of scan. The resulting landmark detection and LV length measurements were displayed and available for immediate evaluation prior to the next image series.

## Results

The trained model was applied to the test datasets. Examples of landmark detection for different long-axis and short-axis views (Fig. 5) demonstrate that the trained model was able to detect the specified landmarks. The detection was robust against the contrast and image appearance difference among three tested imaging sequences. The model was able to perform detection across different short-axis imaging slices, even for the more apical ones where myocardium and RV were different in size and shape, compared to mid-cavity images. For slices above the LV, the model was able to recognize this and not output landmarks.

Table 2 summarizes the detection rate for all views and sequences. For the cine, 99.8% (2,072 out of 2,076) of CH2/CH3/CH4 LAX images and 96.6% (2,906 out of 3008 test images) of SAX images were successfully detected. For the LGE, the detection rates were 99.4% (1,105 out of 1,112 test images) for all LAX views and 97.6% (1,056 out of 1,082 test images) for SAX. For the T1 mapping, the detection rate was 98.9% (439 out of 444 test images).

The few failed detections in LAX test cases were due to incorrect imaging planning, or congenital defects leading to unusual shapes of LV or due to bad image quality. Examples of mis-detected LAX cases and discussion can be found in Appendix E1.

For the 102 mis-detected SAX cases in cine, 51 missed the A-RVI and 25 missed the P-RVI and 13 missed LV center. 50% of error cases were found to be on the most basal and apical slices (defined as top two slices or the last slice for a SAX stack). For the 26 failed SAX cases in LGE, 7 missed the A-RVI and 1 missed the P-RVI and 2 missed LV center. 11 errors were due to unnecessary landmarks detected in slices outside LV. All T1 MOLLI failures (6 out of a total of 444 test cases) missed P-RVI, due to unusual imaging planning for one patient. Examples of mis-detected SAX cases can be found in Appendix E2.

For all cases where detection was successful, the L2 distances between model detection and expert labels were computed. For the SAX, the angle of A-RVI was measured. For the LAX, the LV length was computed. Table 3 summarized the L2 distances and two derived measurements, reported separately for all imaging views and imaging sequences. The L2 distances between the trained model and the 1$^{st}$ operator were between 2 to 3.5 mm. Figure 6 gives detection examples with model derived and manual landmarks and their L2 distance reported, showing model landmarks were in close vicinity to the manual labels. Table 3 listed L2 distances between two operators for the labelled portion of tested data. The L2 distances between two human operators were comparable to model distances. No significant differences were found for the A-RVI angle and LV length measurement between the trained models and the 1$^{st}$ operator for all imaging applications and imaging views. For the test data labelled by both operators, no statistically significant differences were found between two operators for both measures.

The deployed model was tested on the MR scanner for measurement of processing speed. On a tested server (2x Intel Xeon E5-2640 v3@3.400GHz, without GPU), it took ~74ms to load the model and ~5.6s to apply the model on all 30 phases of a cine series on CPU. When tested on a server with GPU (2x Intel Xeon Gold 6152@2.101GHz, 1x NVIDIA RTX 2080Ti), model loading took 66 ms and applying model took 610 ms. Appendix E3 provides more information for this landmark detection application. A movie of this example can be found in Supplemental Data.

## Discussion

This study presents a CNN based solution for landmark detection in cardiac MR images. Three CMR imaging applications, cine, LGE and T1 mapping were tested in this study. A multi-task learning strategy was used to simplify the training and ease deployment. A large dataset was curated from two collaborative hospitals, consisting of 2,329 patients (34,019images) for training and 531 patients (7,723 images) for test. Among the whole training dataset, cine images took the majority (~86%). As a result, a transfer learning strategy with fine tuning was applied to improve the performance of the LGE and T1 mapping detection. The resulting models were robust across different imaging views and imaging sequences. An inline application was built to demonstrate the clinical usage of landmark detection to automatically measure and output LV length on the MR scanner.

Landmark detection using deep learning has not been extensively studied for cardiac MR imaging, but had been investigated for computer vision applications, such as facial key point detection (29,30) or human pose estimation (22,31). In the context of facial point detection, a few open datasets are available, including the past Kaggle Facial Keypoints Detection competition (32). Two categories of approaches were explored for key point detection. First, the output layer of a CNN explicitly computes the x-y coordinates of landmark points and L2 regression loss was used for training. Second, landmark coordinates were implicitly coded as heat-maps. In this context, the detection problem was reformulated as a segmentation problem. In the human pose estimation, the segmentation-based models outperformed regression models (22,33). Here fewer landmarks were detected and were more spatially sparse distributed. The human pose images had much more variation, compared to human faces which often had been pre-processed as front position (34). It is easier for heat-map detection to handle landmark occlusion. For example, in Figure 1, some images may not include targeted landmarks, which is represented by low probability of detection outputs. For these reasons, this study adopted the segmentation model for CMR.

Detection was slightly less accurate on basal and apical imaging SAX slices. In these regions, the "ambiguity" of anatomy increased, leading to more variant in data labelling and more difficulties for model to give correct inference. While different neural network architecture or loss function may be experimented for better accuracy, the limit of accuracy may be on the data labelling. Overall, the models performed better in LAX views than SAX slices. The reason is the less imaging and anatomical variation in long-axis acquisition. For a correctly prescribed LAX imaging slice, occlusion does not happen. A related finding is the detection of mid-cavity SAX slices was very robust. Therefore, future improvement in data labelling shall focus on the basal and apical SAX slices. There are a few failed detections due to unusual anatomy, inferior image quality and bad slide planning. More specific data collection for these "long-tail" scenario is needed to further improve models. One plausible strategy is to deploy models and monitor performance regularly and collect corner cases.

### *Limitations*

There are limitations to this study. First, a single operator labelled entire datasets with significant efforts. Due to the limitation of research resources, the 2$^{nd}$ operator only labeled portion of test set to measure inter-operator variation. Second, three imaging applications were tested in this study. If the model was to be applied to the detection of a new anatomy (e.g. RV center) or a new imaging sequence or a different cardiac view, more training data will be required, but using transfer learning would reduce the amount of new data needed. The development process will have to be iterative to cover more imaging sequences and anatomy.

Third, MR scanners from Siemens were used in this study. A recent study (35) reported performance of deep learning models trained on one vendor may drop for different vendors although augmentation was used to improve robustness. It is very likely to require further data and training to extend current model for MRI from other vendors.

*Conclusion*

In this study, a CNN based solution for landmark detection was developed and validate for cardiac MRI. A large training dataset of 2,329 patients was collected. Test was performed on 531 consecutive patients from two centers. The resulting models were robust across different imaging views and imaging sequences. Quantitative validation showed the CNN detection performance was comparable to the inter-operator variation. Based on the detected landmarks, RV insertion and LV length can be reliably measured.

**Appendix E1: Examples of failed detection for LAX views.**

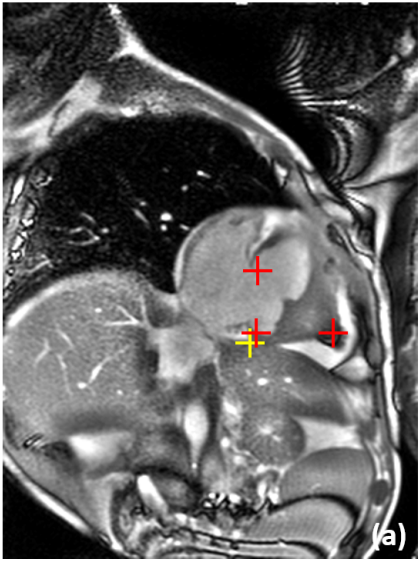
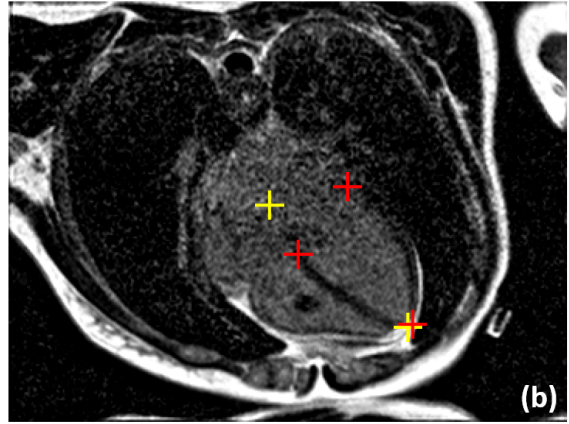
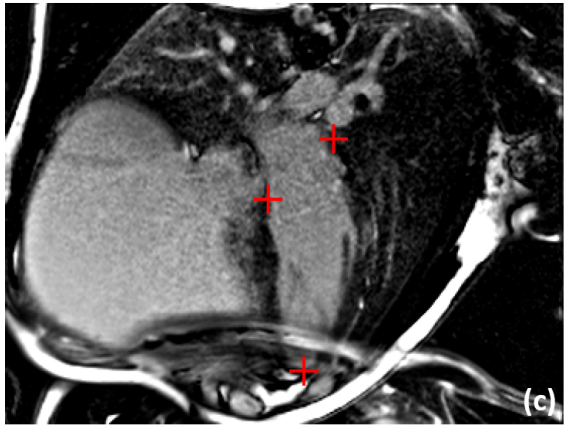
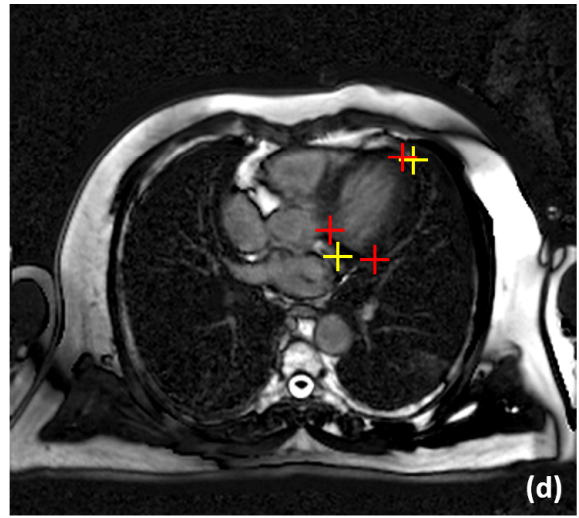

(a) This CH2 cine image contains unusual anatomy, due to congenital heart abnormality of this patient. Model missed two landmarks on this image. (b) This LGE image had very low signal-noise-ratio. The model correctly detected apical point but missed other landmarks. (c) An LGE image had severe aliasing artifacts, causing models to miss all three landmarks. (d) The acquisition plane of this CH4 LGE image was imperfectly placed, causing the model to miss landmarks. +: manual landmarks; +: model landmarks

**Appendix E2: Examples of failed detection for SAX views.**

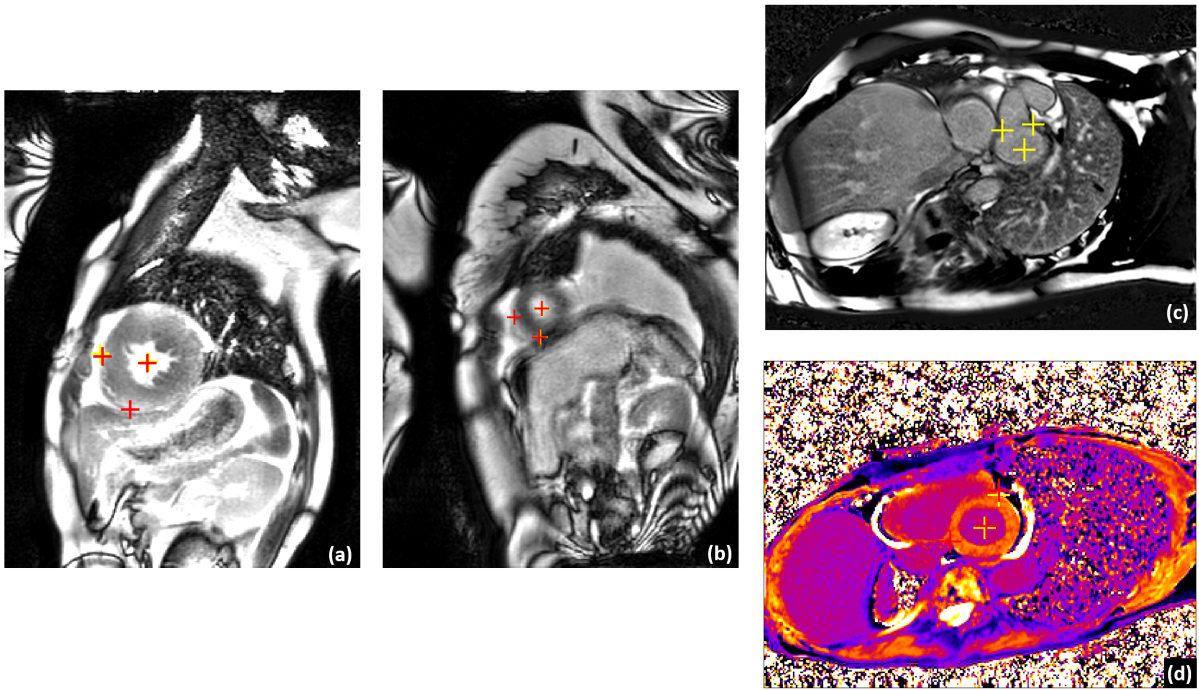

(a) Detection failed to find the P-RVI point on this cine image, due to the very small RV cavity. (b) Both RVI points were missed in this very apical cine slice. (c) This LGE image was acquired outside the LV, but model incorrectly outputted landmarks. (d) The P-RVI point was missed in this pre-contrast T1 map, likely due to non-standard imaging plane subscribed.
+: manual landmarks; +: model landmarks

**Appendix E3: Model deployment and scanner integration.**

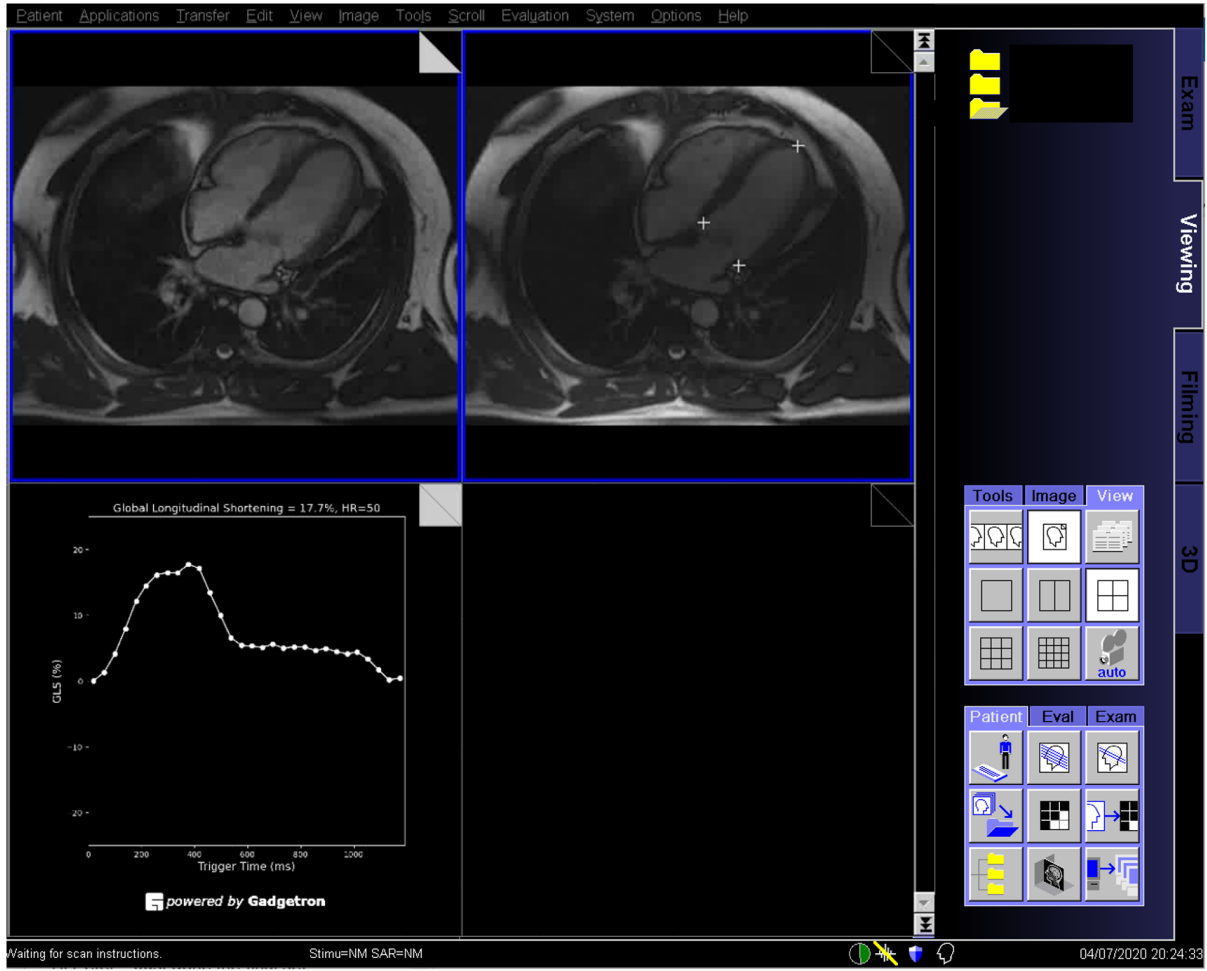

The trained landmark detection models can be useful for many CMR analysis tasks. As an example, the model for LAX detection was integrated on MR scanner and used to measure LV length for long-axis cine image series. The global longitudinal shortening ratio can be computed from the AI measurement as:

$$100 \times (LV\_length_{ED} - LV\_length_{ES})/LV\_length_{ED}$$

In this example, a scanner screen snapshot shows a four-chamber cine processed with proposed landmark detection algorithm. The LV length for every cardiac phase was measured and longitudinal shortening ratio was computed. This approach was fully automated. The corresponding movie of this example can be found in the supplement data or at https://www.youtube.com/watch?v=Xekpfx82gWk.

**Availability of data and material**

The raw data that support the findings of this study are available from the corresponding author upon reasonable request subject to restriction on use by the Office of Human Subjects Research. The source file to train the CNN model and example datasets are shared at https://github.com/xueh2/CMR_LandMark_Detection.git

**Funding**

Supported by the National Heart, Lung and Blood Institute, National Institutes of Health by the Division of Intramural Research.

**Authors' contributions**

HX and PK conceived of the study and drafted the manuscript. HX and PK developed the algorithms, implemented the inline integration of neural net model and performed processing and analysis. HX and JA labelled the datasets. RD, MF, JM performed the patient studies used to acquire training data. All authors participated in revising the manuscript and read and approved the final manuscript.

# References

1. Kellman P, Hansen MS. T1-mapping in the heart: accuracy and precision. J Cardiovasc Magn Reson. 2014;16:2.

2. Giri S, Chung Y-C, Merchant A, et al. T2 quantification for improved detection of myocardial edema. J Cardiovasc Magn Reson. 2009;11:56.

3. Kellman P, Hernando D, Arai AE. Myocardial Fat Imaging. Curr Cardiovasc Imaging Rep. 2010;3(2):83–91.

4. Xue H, Brown LAE, Nielles-Vallespin S, Plein S, Kellman P. Automatic In-line Quantitative Myocardial Perfusion Mapping : processing algorithm and implementation. Magn Reson Med. 2019;83(2):712–730.

5. Kellman P, Hansen MS, Nielles-Vallespin S, et al. Myocardial perfusion cardiovascular magnetic resonance: optimized dual sequence and reconstruction for quantification. J Cardiovasc Magn Reson. 2017;19(1):43.

6. Nickander J, Themudo R, Sigfridsson A, Xue H, Kellman P, Ugander M. Females have higher myocardial perfusion, blood volume and extracellular volume compared to males – an adenosine stress cardiovascular magnetic resonance study. Sci Rep. Springer US; 2020;10(1).

7. Schulz-Menger J, Bluemke DA, Bremerich J, et al. Standardized image interpretation and post-processing in cardiovascular magnetic resonance - 2020 update : Society for Cardiovascular Magnetic Resonance (SCMR): Board of Trustees Task Force on Standardized Post-Processing. J Cardiovasc Magn Reson. Journal of Cardiovascular Magnetic Resonance; 2020;22(1):19.

8. Bhuva AN, Bai W, Lau C, et al. A Multicenter, Scan-Rescan, Human and Machine Learning CMR Study to Test Generalizability and Precision in Imaging Biomarker Analysis. Circ Cardiovasc Imaging. 2019;12(10):1–11.

9. Bai W, Sinclair M, Tarroni G, et al. Automated cardiovascular magnetic resonance image analysis with fully convolutional networks. J Cardiovasc Magn Reson. Journal of Cardiovascular Magnetic Resonance; 2018;20(65).

10. Xue H, Davies R, Brown LA, et al. Automated Inline Analysis of Myocardial Perfusion MRI with Deep Learning. Radiol Artif Intell. In Press.

11. Knott KD, Seraphim A, Augusto JB, et al. The Prognostic Significance of Quantitative Myocardial Perfusion: An Artificial Intelligence Based Approach Using Perfusion Mapping. Circulation. 2020;1282–1291.

12. Asgeirsson D, Hedström E, Jögi J, et al. Longitudinal shortening remains the principal component of left ventricular pumping in patients with chronic myocardial infarction even when the absolute atrioventricular plane displacement is decreased. BMC Cardiovasc Disord. BMC Cardiovascular Disorders; 2017;17(1):1–9.

13. Messroghli DR, Walters K, Plein S, et al. Myocardial T1 mapping: application to patients with acute and chronic myocardial infarction. Magn Reson Med.22

**Table 1. Information for training and test dataset distribution and acquisition.**

| | Imaging | View | #Patients | #Images | #Scan duration |
|---|---|---|---|---|---|
| **Training** | Cine | CH2 | 2,115 | 4,232 | 2017/12/18 to 2017/12/29<br>2018/01/02 to 2018/01/28<br>2020/01/02 to 2020/04/19 |
| | | CH3 | 2,102 | 4,206 | |
| | | CH4 | 2,127 | 4,256 | |
| | | SAX | 702 | 16,520 | |
| | LGE | CH2 | 599 | 599 | 2020/01/02 to 2020/02/29 |
| | | CH3 | 582 | 582 | |
| | | CH4 | 599 | 599 | |
| | | SAX | 178 | 2,018 | |
| | T1 MOLLI | SAX | 202 | 1,077 | 2020/01/02 to 2020/03/25 |
| | **Total** | **ALL** | **2,329** | **34,019** | *three consecutive periods* |
| | Imaging | View | #Patients | #Images | #Scan duration |
| **Testing** | Cine | CH2 | 347 | 694 | 2020/05/01 to 2020/07/03 |
| | | CH3 | 345 | 690 | |
| | | CH4 | 347 | 692 | |
| | | SAX | 128 | 3,008 | |
| | LGE | CH2 | 370 | 370 | 2020/05/01 to 2020/07/03 |
| | | CH3 | 370 | 370 | |
| | | CH4 | 370 | 372 | |
| | | SAX | 96 | 1,082 | |
| | T1 MOLLI | SAX | 161 | 445 | 2020/05/01 to 2020/07/23 |
| | **Total** | **ALL** | **531** | **7,723** | 2020/05/01 to 2020/07/23 |

Table 2. Detection rate for three imaging applications at all tested CMR views.

| Imaging | View | #Tested Images | #Success | Detection rate |
|---|---|---|---|---|
| Cine | CH2 | 694 | 692 | 99.7% |
| | CH3 | 690 | 688 | 99.7% |
| | CH4 | 692 | 692 | 100% |
| | SAX | 3,008 | 2,906 | 96.6% |
| LGE | CH2 | 370 | 368 | 99.5% |
| | CH3 | 370 | 368 | 99.5% |
| | CH4 | 372 | 369 | 99.2% |
| | SAX | 1,082 | 1,056 | 97.6% |
| T1 MOLLI | SAX | 445 | 439 | 98.9% |

Table 3. Summary of quantitative evaluation of CMR landmark detection for successfully detected cases. "1st vs. AI" is to compare the manual labels of the first operator to the trained model. "1st vs. 2nd" is to compare two human operators for the portion of test data labelled by both.

| Imaging | View | Landmark | L2 distance | | LV length difference in % | |
|---|---|---|---|---|---|---|
| | | | 1st vs AI | 1st vs. 2nd | 1st vs AI | 1st vs. 2nd |
| Cine | CH2 | A-P | 2.1±1.8 | 2.8±1.9 | 2.0±1.7, P=0.42 | 1.9±1.4, P=0.95 |
| | | I-P | 2.4±2.0 | 3.0±3.9 | | |
| | | APEX* | 2.4±1.8 | 4.1±2.8 | | |
| | CH3 | IL-P | 2.4±1.7 | 2.8±1.6 | 1.5±1.3, P=0.79 | 2.0±1.7, P=0.97 |
| | | AS-P* | 2.2±1.5 | 4.0±2.4 | | |
| | | APEX* | 3.2±2.4 | 3.8±2.1 | | |
| | CH4 | AL-P | 3.4±2.1 | 3.5±2.0 | 1.4±1.2, P=0.92 | 2.0±1.4, P=0.77 |
| | | IS-P* | 2.1±1.7 | 2.6±1.6 | | |
| | | APEX | 2.8±1.9 | 2.8±1.6 | | |
| LGE | CH2 | A-P | 2.9±2.6 | 3.3±2.0 | 2.7±2.5, P=0.16 | 2.5±2.1, P=0.82 |
| | | I-P | 3.4±2.7 | 3.4±2.5 | | |
| | | APEX | 3.1±2.6 | 3.4±2.5 | | |
| | CH3 | IL-P | 3.4±3.1 | 3.5±2.1 | 2.6±2.6, P=0.37 | 2.9±2.2, P=0.34 |
| | | AS-P* | 2.7±2.1 | 3.6±2.3 | | |
| | | APEX | 3.3±2.8 | 3.3±2.5 | | |
| | CH4 | AL-P | 3.1±1.6 | 3.3±2.2 | 2.0±1.4, P=0.13 | 1.9±1.9, P=0.53 |
| | | IS-P | 2.0±1.5 | 2.5±2.3 | | |
| | | APEX | 2.7±1.2 | 2.1±1.6 | | |

| Imaging | View | Landmark | L2 distance | | A-RVI angle difference in degree | |
|---|---|---|---|---|---|---|
| | | | 1st vs AI | 1st vs. 2nd | 1st - AI | 1st - 2nd |
| Cine | SAX | A-RVI | 3.1±1.8 | 3.5±2.6 | 1.3±3.4, P=0.14 | -0.7±4.1, P=0.89 |
| | | P-RVI | 2.4±2.1 | 2.7±1.6 | | |
| | | C-LV | 2.0±1.1 | 2.4±1.2 | | |
| LGE | SAX | A-RVI | 3.0±3.2 | 3.6±3.1 | 0.14±2.9, P=0.92 | -2.0±4.5, P=0.62 |
| | | P-RVI | 2.8±2.6 | 3.3±2.6 | | |
| | | C-LV* | 1.5±0.9 | 2.3±1.1 | | |
| T1 MOLLI | SAX | A-RVI | 2.5±2.0 | 3.0±2.8 | 1.6±3.1, P=0.31 | 1.7±3.9, P=0.41 |
| | | P-RVI | 2.5±2.6 | 2.5±2.0 | | |
| | | C-LV | 1.6±1.0 | 2.0±1.1 | | |

* indicates P<0.05 for the L2 distances between the "1st vs. AI" and "1st vs. 2nd".

# List of Captions

**Figure 1**. Example of CMR images with landmarks. Three short-axis (SAX) views are shown on the top row. The first three images at the 2nd row gives example of long axis views for two chamber (CH2), three chambers (CH3) and four chamber (CH4). For every SAX view, the right ventricular insertion points (1 and 2) are marked, together with center of left ventricular. For the long-axis views, the mitral valve plane points and apical point are marked. Note for some SAX slices (the rightmost column), no landmarks can be identified. The last column gives examples of LGE images and T1 maps. Transfer learning was applied to detect landmarks from these imaging applications.

**Figure 2**. Example of surface coil inhomogeneity correction. A two-chamber slice before and after correcting the surface coil inhomogeneity. In this study, a copy of original image (left) and its corrected version (right) was kept in the dataset and randomly picked to feed into the CNN as a data augmentation step.

**Figure 3**. The landmark detection problem can be reformulated as a semantic segmentation problem. Every landmark point in this CH2 image on the left can be convolved with a gaussian kernel and converted into a spatial probability map or heat map (upper row, from left to right, probability for background, anterior valve point, inferior valve point and apex). Unlike the binary detection task with target being one-hot binary mask, loss functions working on continues probability such as the KL divergence are needed.

**Figure 4**. The backbone CNN network developed for landmark detection has a U-net structure. More layers can be inserted to both downsampling and upsampling branches. More blocks can be inserted into each layer. The output layer outputs the per-pixel scores which goes through Softmax function. For the LAX detection, data from three views were trained together for one model. As shown in the input, every minibatch was assembled by randomly selected images from three views and used for back propagation. A total of four layers were used in this experiment with 3 or 4 blocks per layer. The output tensor shapes were reported in the figure, in the format of [B, C, H, W]. B is the size of minibatch and C is the number of channels. Input images have one channel for image intensity and output has four channels for three landmarks and background. The illustration here for outputs plots three landmark channels color-coded and omits the background channel.

**Figure 5**. Examples of landmark detection. The left panel are cine detection examples for long and short-axis images. The right panel are LGE and T1 examples, where the first two rows are examples of LGE images with different imaging slices. The third and last row are T1 mapping detection.

**Figure 6**. Examples of L2 distance of landmarks. For every pair of manual and model delineated landmarks, the L2 distance (mm) is labeled. Red: manual landmarks; Yellow: model landmarks.

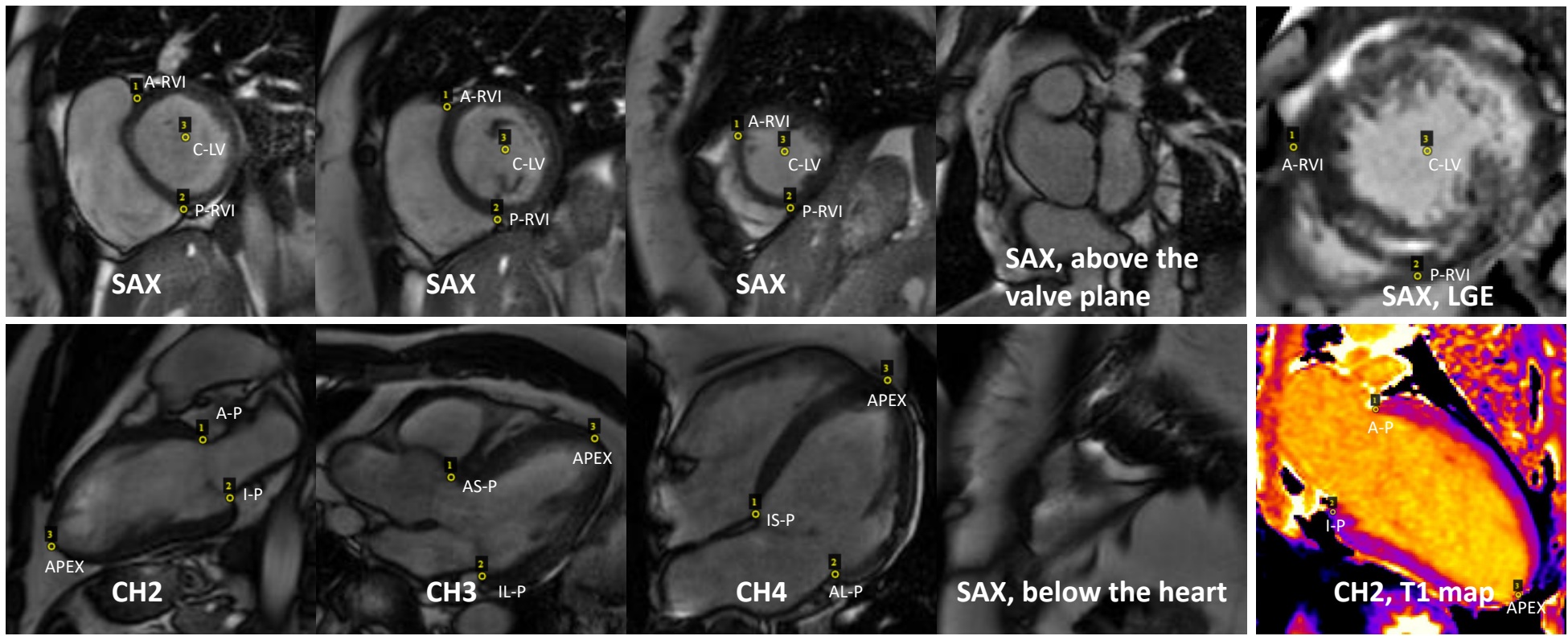

Figure 1. Example of CMR images with landmarks. Three short-axis (SAX) views are shown on the top row. The first three images at the 2nd row gives example of long axis views for two chamber (CH2), three chambers (CH3) and four chamber (CH4). For every SAX view, the right ventricular insertion points (1 and 2) are marked, together with center of left ventricular. For the long-axis views, the mitral valve plane points and apical point are marked. Note for some SAX slices (the rightmost column), no landmarks can be identified. The last column gives examples of LGE images and T1 maps. Transfer learning was applied to detect landmarks from these imaging applications.

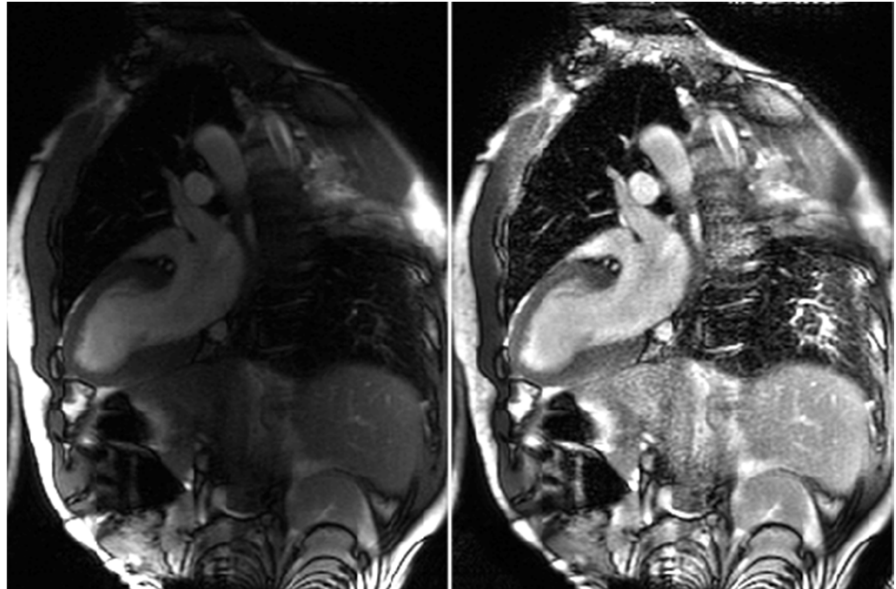

Figure 2. Example of surface coil inhomogeneity correction. A two-chamber slice before and after correcting the surface coil inhomogeneity. In this study, a copy of original image (left) and its corrected version (right) was kept in the dataset and randomly picked to feed into the CNN as a data augmentation step.

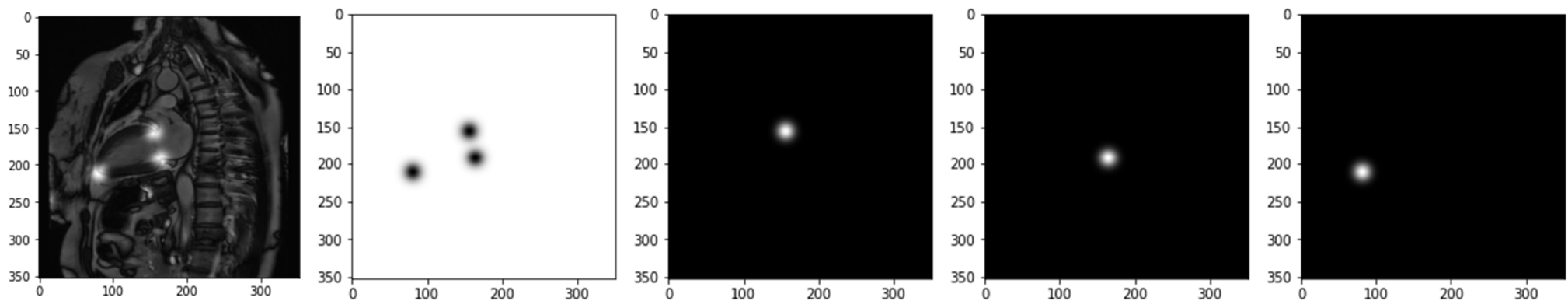

Figure 3. The landmark detection problem can be reformulated as a semantic segmentation problem. Every landmark point in this CH2 image on the left can be convolved with a gaussian kernel and converted into a spatial probability map or heat map (upper row, from left to right, probability for background, anterior valve point, inferior valve point and apex). Unlike the binary detection task with target being one-hot binary mask, loss functions working on continues probability such as the KL divergence are needed.

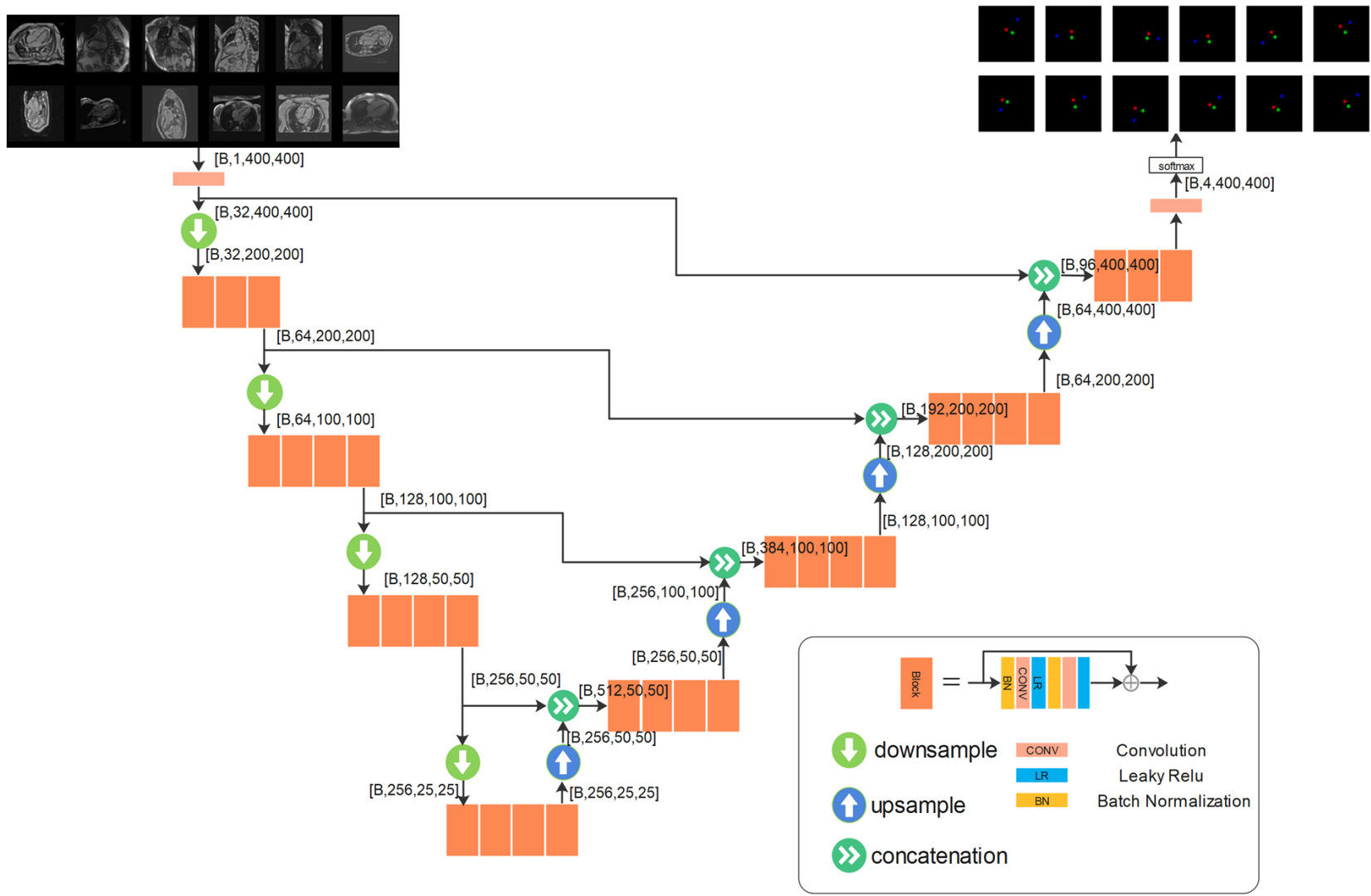

Figure 4. The backbone CNN network developed for landmark detection has a U-net structure. More layers can be inserted to both downsampling and upsampling branches. More blocks can be inserted into each layer. The output layer outputs the per-pixel scores which goes through Softmax function. For the LAX detection, data from three views were trained together for one model. As shown in the input, every minibatch was assembled by randomly selected images from three views and used for back propagation. A total of four layers were used in this experiment with 3 or 4 blocks per layer. The output tensor shapes were reported in the figure, in the format of [B, C, H, W]. B is the size of minibatch and C is the number of channels. Input images have one channel for image intensity and output has four channels for three landmarks and background. The illustration here for outputs plots three landmark channels color-coded and omits the background channel.

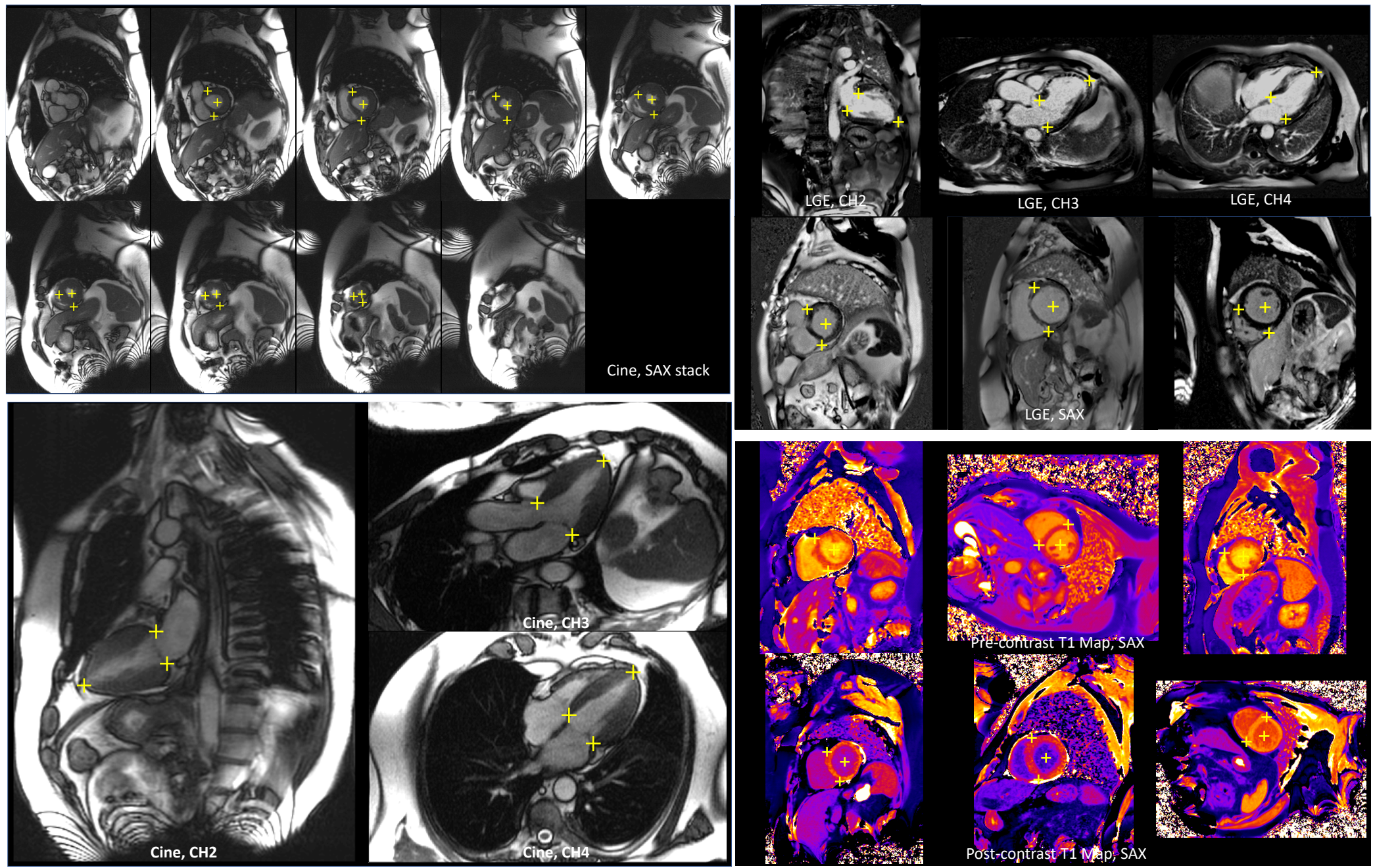

Figure 5. Examples of landmark detection. The left panel are cine detection examples for long and short-axis images. The right panel are LGE and T1 examples, where the first two rows are examples of LGE images with different imaging slices. The third and last row are T1 mapping detection.

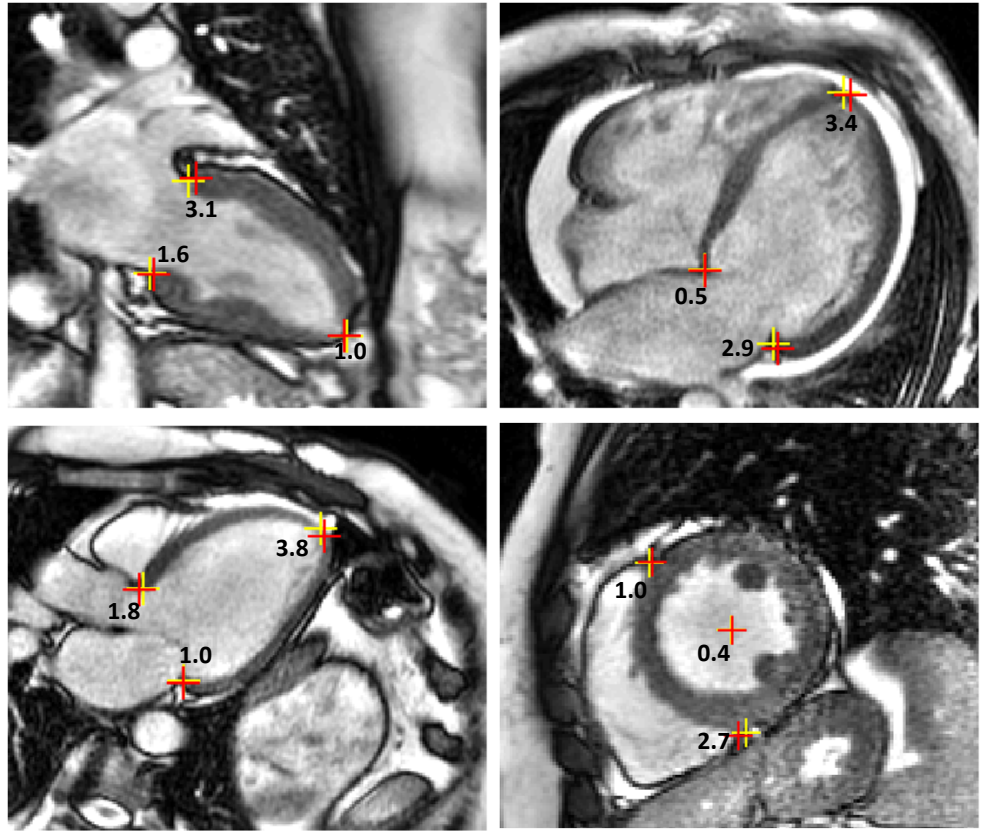

Figure 6. Examples of L2 distance of landmarks. For every pair of manual and model delineated landmarks, the L2 distance (mm) is labeled.
+: manual landmarks; +: model landmarks